\input{aipcheck}

\documentclass[
    ,final            
  ]
  {aipproc}

\layoutstyle{6x9}

\newcommand{\ba}{\begin{eqnarray}}
\newcommand{\ea}{\end{eqnarray}}

\newcommand{\DD}{{\cal {D}}}

\newcommand{\bben}{\begin{itemize}}
\newcommand{\eeen}{\end{itemize}}
\newcommand{\bbq}{\begin{quote}}
\newcommand{\eeq}{\end{quote}}

\newcommand{\RR}{{}^3{\cal{R}}}

\newcommand{\T}{{\cal{T}}^{(3)}}

\newcommand{\EE}{{\cal{E}}}

\newcommand{\JJ}{{\cal{J}}}

\newcommand{\HH}{{\cal{H}}}

\newcommand{\PP}{{\cal{P}}}
\newcommand{\MM}{{\cal{M}}}

\newcommand{\Da}{\delta^{(A)}}

\newcommand{\Dth}{\delta^{(\Theta)}}
\newcommand{\dDth}{\dot\delta^{(\Theta)}}
\newcommand{\Dm}{\delta^{(\mu)}}
\newcommand{\dDm}{\dot\delta^{(\mu)}}

\newcommand{\Dp}{\delta^{(p)}}

\newcommand{\dd}{{\rm{d}}}

\begin{document}

\title[Quasi--local variables and inhomogeneous cosmological sources]{Quasi--local variables and inhomogeneous cosmological sources with spherical symmetry.}


\author{Roberto A. Sussman}{
  address={Instituto de Ciencias Nucleares, UNAM, M\'exico D.F. 04510, M\'exico}
}



\begin{abstract}
 We examine a large class of inhomogeneous spherically symmetric spacetimes that generalize the Lema\^\i tre--Tolman--Bondi dust solutions to nonzero pressure (``LTB spacetimes''). Local covariant LTB objects can be expressed as perturbations of covariant quasi--local (QL) scalars that satisfy evolution equations of equivalent Friedman--Lema\^\i tre-Robertson--Walker (FLRW) scalars. Thus, the dynamics of these spacetimes can be rigorously described as non--linear, gauge invariant and covariant perturbations on a formal FLRW background given by the QL scalars. Since LTB spacetimes are compatible with a wide variety of ``equations of state'' and theoretical assumptions, they provide an ideal framework for numerical models of cosmological sources under idealized but fully non--linear conditions. As an illustrative example, we briefly examine the formation of a black hole in an expanding Chaplygin gas universe.       
\end{abstract}

\maketitle


\section{Introduction.}

 The dominant theoretical explanation to account for the cosmic  acceleration detected in recent observations is based on an elusive source, ``dark energy'' (DE) that behaves as a cosmological constant or as a fluid with negative pressure. While dark matter (DM) in galactic halos is assumed to be inhomogeneous (and Newtonian) at the galactic scale, dark energy is usually studied by means of FLRW models and/or their linear perturbations. The justification is that the dynamical effects of DE are relevant only in larger scales (100-300 Mpc), in which the universe appears to be homogeneous or nearly so~\cite{review}. However, as long as the fundamental nature of DM and DE is not known, there is no reason to assume {\it a priori} that no new valuable information could come from studying these sources and their interactions under  inhomogeneous, relativistic and fully non--linear conditions, at the very least in the galactic and intermediate large scale. 
  
Since fully general inhomogeneity requires numerical codes of high complexity, we offer in this article a compromise by looking at spherically symmetric sources, which are obviously more idealized but still useful to examine non--linear phenomena that cannot be studied with linear perturbations. The resulting models, ``LTB spacetimes'', can be fully described by autonomous first order evolution equations that can be well handled by simple numerical methods. These models are quite general and readily allow for an inhomogeneous generalization of a large number of known FLRW solutions.  See \cite{sussQLnum,sussQL} for a more comprehensive discussion of the contents of this article.    
  
\section{LTB spacetimes in the ``fluid flow'' description.}

Spherically symmetric inhomogeneous dust sources are usually described by the well known Lema\^\i tre--Tolman--Bondi metric~\cite{kras,ltbstuff,suss02}
\begin{equation}\dd s^2 = -c^2\dd t^2+ \frac{R'{}^2}{1-K}\dd r^2+R^2\left(\dd\theta^2+\sin^2\theta\dd\phi^2\right).\label{ltb}\end{equation}
where $R=R(ct,r)$,\, $R'=\partial R/\partial r$ and $K=K(r)$. A large class of spherically symmetric spaectimes follow at once by considering the most general source for (\ref{ltb}) in a comoving frame ($u^a=\delta^a_0$), which is the energy--momentum tensor
\begin{equation}T^{ab}=\mu\,u^au^b+p\,h^{ab}+\Pi^{ab}, \label{Tab}\end{equation}
where $\mu$ and $p$ are the matter--energy density and the isotropic pressure, $h^{ab}=u^au^b+g^{ab}$ is the induced metric of hypersurfaces $\T$ orthogonal to $u^a$, and $\Pi^{ab}$ is the symmetric traceless tensor of anisotropic pressure. We will call ``LTB' spacetimes'' to all solutions of Einstein's equations for (\ref{ltb}) and (\ref{Tab}). 

Besides the scalars $\mu$ and $p$, and the tensor $\Pi^{ab}$, the remaining basic covariant objects of LTB spacetimes are: 
\ba \Theta &=& \tilde\nabla_au^a=\frac{2\dot R}{R}+\frac{\dot R'}{R'},\qquad \hbox{Expansion scalar}\label{Theta}\\
\RR &=& \frac{2(KR)'}{R^2R'},\hskip 2.2cm \hbox{ Ricci scalar of the hypersurfaces}\,\, \T,\label{RR}\\
\sigma_{ab} &=& \tilde\nabla_{(a}u_{b)}-\frac{\Theta}{3}h_{ab}\qquad\quad \hbox{ Shear tensor},\label{shear}\\
E^{ab} &=& u_cu_d C^{abcd}\hskip 2.2cm  \hbox{Electric Weyl tensor}\label{EWeyl}\ea
where $\dot R=u^a\nabla_a R$,\, $\tilde\nabla_a = h_a^b\nabla_b$,\, and $C^{abcd}$ is the Weyl tensor. For spherically symmetric spacetimes, the symmetric traceless tensors $\sigma^{ab},\,\Pi^{ab}$ and $E^{ab}$ can be expressed in terms of  single scalar functions as
\begin{equation} \sigma^{ab}=\Sigma\,\Xi^{ab},\qquad \Pi^{ab}=\PP\,\Xi^{ab},\qquad E^{ab}=\EE\,\Xi^{ab},  \label{PSEsc}\end{equation}
where $\Xi^{ab}=h^{ab}-3\eta^a\eta^b$ and $\eta^a=\sqrt{h^{rr}}\delta^a_r$ is the unit vector orthogonal to $u^a$ and to the 2--spheres orbits of SO(3) parametrized by $(\theta,\phi)$. The field equations  $G^{ab}=\kappa T^{ab}$ (with $\kappa=8\pi G/c^4$) for (\ref{ltb}) and (\ref{Tab}) are
\ba \kappa\,\mu\,R^2R' &=& \left[R(\dot R^2+K)\right]',\label{mu1}\\
\kappa\,p\,R^2R' &=& -\frac{1}{3}\left[R(\dot R^2+K)+2R^2\ddot R\right]',\label{p1}\\
\kappa\,\PP\,\frac{R'}{R} &=& -\frac{1}{6}\left[\frac{\dot R^2+K}{R^2} +\frac{2\ddot
Y}{Y}\right]',\label{PP1}\ea
while from (\ref{Theta}) and (\ref{PSEsc}), we obtain for $\EE$ and $\Sigma$ 
\begin{equation}\Sigma = \frac{1}{3}\left[\frac{\dot R}{R}-\frac{\dot R'}{R'}\right],\qquad
\EE = -\frac{\kappa}{2}\,\PP-\frac{\kappa}{6}\,\mu+ \frac{\dot
R^2+K}{2R^2}.\label{SigEE1}\end{equation}
Bearing in mind (\ref{PSEsc}), all covariant objects (scalars and proper tensors) in LTB spacetimes can be fully characterized by the local covariant scalars $\{\mu,\,p,\,\PP,\,\Theta,\,\Sigma,\,\EE,\,\RR\}$. Given the covariant ``1+3'' slicing afforded by $u^a$,  their evolution can be completely determined by a ``fluid flow''  description of scalar evolution equations for these scalars (as in \cite{ellisbruni89,BDE,1plus3}). However, we will consider another covariant scalar representation. 

\section{Quasi--local (QL) variables}

The Misner--Sharp quasi--local mass--energy function, $\MM$, is a well known invariant in spherically symmetric spacetimes~\cite{MSQLM,kodama,szab,hayward1}. For LTB spacetimes (\ref{ltb})--(\ref{Tab}) it satisfies the equations
\ba 2\MM' &=&\kappa \mu\,R^2R',\label{Mr}\\ 
2\dot\MM &=& -\kappa\,(p-2\PP)\,R^2 \dot R.\label{Mt}  \ea
Comparing (\ref{Mr}) with the field equation (\ref{mu1}) suggest obtaining an integral expression for $\MM$ that can be related to $R$ and $\dot R$. This integral along the $\T$ exists and is bounded if the integration domain contains a symmetry center \cite{hayward1}. Assuming as integration domain a spherical comoving region $\DD=\xi\times{\bf\rm{S}}^2\subset \T$, where ${\bf\rm{S}}^2$ is the unit 2--sphere, $\xi=\{x\in {\bf\rm{R}}\,|\,0\leq x\leq r\}$ and $x=0$ marks a symmetry center so that $\MM(ct,0)=0$ for all $t$, we integrate both sides of (\ref{mu1}) and also (\ref{Mr}). This allows us to define a scalar $\mu_*$ as
\begin{equation} \frac{\kappa}{3}\mu_*\equiv \frac{2\MM}{R^3} =\frac{\kappa}{3}\, \frac{\int_0^r{\mu R^2 R'\dd x}}{\int_0^r{R^2 R'\dd x}}=\frac{\dot R^2+K}{R^2},\label{QLmu}\end{equation}
where $\int_0^r{..\,\dd x}=\int_{x=0}^{x=r}{..\,\dd x}$.  This integral definition of $\mu_*$, which is related to $\mu$ and to the quasi--local mass--energy function, $\MM$, motivates us to define the following:\\

\noindent 
 {\underline{Quasi--local (QL) scalar map}} Let $X(\DD)$ be the set of all smooth integrable scalar functions in $\DD$. For every $A\in X(\DD)$, the quasi--local map is defined as
\begin{equation}\JJ_*:X(\DD)\to X(\DD),\qquad A_*=\JJ_*(A)=\frac{\int_0^r{A R^2 R'\dd x}}{\int_0^r{R^2 R'\dd x}}.\label{QLmap}\end{equation}
The scalar functions $A_*:\DD\to {\bf\rm{R}}$ that are images of $\JJ_*$ will be denoted by ``quasi--local'' (QL) scalars. In particular, we will call $A_*$ the QL dual of $A$. \\

\noindent
Applying the map (\ref{QLmap}) to the scalars $\Theta$ and $\RR$ in (\ref{Theta}) and (\ref{RR}) we obtain
\begin{equation} \Theta_* =\frac{3\dot R}{R},\qquad \RR_* =\frac{6K}{R^2}.\label{QL_TR}\end{equation}
Applying now (\ref{QLmap}) to $\mu$ and $p$, comparing with (\ref{mu1})--(\ref{p1}), and using (\ref{QL_TR}), these two field equations transform into 
\ba \left(\frac{\Theta_*}{3}\right)^2 = \frac{\kappa}{3}\mu_* -\frac{\RR_*}{6},\label{QLfried}\\
\dot \Theta_*= -\frac{\Theta_*^2}{3}-\frac{\kappa}{2}\left(\mu_*+3p_*\right).\label{QLraych}\ea
which are identical to the FLRW Friedman and Raychaudhuri equations, but among QL scalars. These equations can be further combined to yield identically the FLRW energy balance equation:
\begin{equation}\dot\mu_* = -\left(\mu_*+p_*\right)\,\Theta_*.\label{QLebal}\end{equation}
We have found the QL duals for the scalars $\{\mu,p,\Theta,\RR\}$, with the help of (\ref{QLmap}) the remaining covariant scalars $\{\Sigma,\,\PP,\,\EE\}$ can be expressed as deviations or fluctuations of $\mu,\,p$ and $\Theta$ with respect to their QL duals:
\ba \Sigma &=& -\frac{1}{3}\,\left[\Theta-\Theta_*\right],\label{Sigma2}\\
\PP &=& \frac{1}{2}\,\left[p-p_*\right],\label{PP2}\\
\EE &=& -\frac{\kappa}{6}\,\left[\mu-\mu_* +\frac{3}{2}(p-p_*)\right],\label{EE2}\ea
while (\ref{Mt}) becomes
\begin{equation}2\dot\MM=-\kappa\,p_*\,R^2\,\dot R =-\frac{\kappa}{3}p_* \Theta_* R^3.\label{QLp}\end{equation}

\section{Evolution equations for the quasi--local scalars.}\label{eveqs}

The scalars $A$ and $A_*$ are related by the ``relative deviations'' or ``perturbations''
\begin{equation} \Da \equiv \frac{A-A_*}{A_*},\quad \Rightarrow\quad A = A_*\,\left[1+\Da\right].\label{Da_def}\end{equation}
which leads to an alternative QL scalar representation $\{A_*,\,\Da\}$ that it is fully equivalent to the local representation $\{\mu,p,\Theta,\RR,\Sigma,\PP,\EE\}$. Hence, LTB spacetimes are fully determined by evolution equations for the QL scalars.   

It is straightforward to show from (\ref{QLmap}) that the radial gradients of $\mu_*,\,p_*$ and $\HH_*$ can be given in terms of the $\delta$ functions by
\begin{equation} \frac{\Theta_*{}'}{\Theta_*} = \frac{3R'}{R}\,\Dth,\qquad \frac{\mu_*{}'}{\mu_*} = \frac{3R'}{R}\Dm,\qquad \frac{p_*{}'}{p_*} = \frac{3R'}{R}\Dp, \label{rad_grads}\end{equation}
while (\ref{QLraych}) and (\ref{QLebal}) are evolution equations for $\dot\mu_*$ and $\dot\Theta_*$. Hence, the evolution equations for $\Dm$ and $\Dth$ follow from the consistency condition:\,\,
$\left[A_*{}'\right]\,\dot{}=\left[\dot A_*\right]'$,
applied to (\ref{QLraych}), (\ref{QLebal}) and (\ref{rad_grads}) for $A_*=\Theta_*,\,\mu_*$. The result is the following set of autonomous evolution equations for the QL scalar representation $\{A_*,\,\Da\}$:
\ba 
\dot\mu_* &=& -\left[\,1+w\,\right]\,\mu_*\,\Theta_*,\label{evmu_ql}\\
\dot\Theta_* &=& -\frac{\Theta_*^2}{3} -\frac{\kappa}{2}\,\left[\,1+3\,w\,\right]\,\mu_*,
\label{evHH_ql}\\
\dDm &=& \Theta_*\,\left[\left(\Dm-\Dp\right)\,w-\left(1+w+\Dm\right)\Dth\right],
\label{evDmu_ql}\\
\dDth &=& -\frac{\Theta_*}{3}\,\left(1+\Dth\right)\,\Dth + \frac{\kappa\mu_*}{6\,(\Theta_*/3)}\left[\Dth-\Dm+3w\,\left(\Dth-\Dp\right)\right],
\label{evDth_ql}
\ea
where
\begin{equation}w\equiv \frac{p_*}{\mu_*}.\label{wdef}\end{equation}
The constraints associated with these evolution equations are simply the spatial  gradients (\ref{rad_grads}), while the Friedman equation (or Hamiltonian constraint) is (\ref{QLfried}). Notice that the constraints (\ref{rad_grads}) follow directly from differentiating the integral definition (\ref{QLmap}), so by using the QL variables we do not need to solve these constraints in order to integrate (\ref{evmu_ql})--(\ref{evDth_ql}).  It is straightforwards to  prove (see \cite{sussQL}) that the evolution equations (\ref{evmu_ql})--(\ref{evDth_ql}) and the constraints (\ref{QLfried}) and (\ref{rad_grads}) are wholly equivalent to the 1+3 evolution equations and constraints for LTB models in the ``fluid flow'' description of Ellis, Bruni, Dunsbury and van Ellst \cite{ellisbruni89,BDE,1plus3}.    

\section{A non--linear perturbation scheme}\label{perturb}

The definition (\ref{Da_def}) and the evolution equations (\ref{evmu_ql})--(\ref{evDth_ql}) suggest that $\Da$ can be rigorously defined as  spherical perturbations on a formal FLRW ``background'' state given by the $A_*$.  Considering the perturbation formalisms developed by  Ellis, Bruni and Dunsbury~\cite{ellisbruni89,BDE,1plus3} and Bardeen~\cite{bardeen}, a perturbation scheme based on the $\Da$ can be defined rigorously for the FLRW-LTB case in terms of a suitable map between $\bar X$ and $X$, which are, respectively, the sets of smooth integrable scalar functions in $\bar S$ (FLRW model) and $S$ (the ``perturbed'' lumpy LTB model). For all covariant FLRW  scalars $\bar A\in \bar X$ (we denote FLRW objects with an over--bar) this map is
\begin{equation} \Phi: \bar X\to X,\qquad \Phi(\bar A)=\JJ_*(A)=A_*\in X,\quad\Rightarrow\quad \Da = \frac{A-\Phi(\bar A)}{\Phi(\bar A)}\label{Phi}\end{equation}
and characterizes QL scalars (which are LTB objects satisfying FLRW dynamics) as the ``background model'' in LTB spacetimes. Following Dunsbury, Ellis and Bruni~\cite{ellisbruni89,BDE}, a perturbation scheme on FLRW cosmologies is covariant if $S$ is described by the ``1+3'' fluid flow variables of\cite{ellisbruni89,BDE,1plus3}. Although our description of LTB spacetimes is not based on these scalars (the local covariant scalars), it is still covariant because $\mu_*,\,p_*$ and $\Theta_*$, are themselves covariant scalars by virtue of their connection with the invariants $\MM,\,R$ and their derivatives in (\ref{QLmu}),  (\ref{QL_TR}) and (\ref{QLp}) (see \cite{kodama}). Hence, the formalism associated with (\ref{Phi}) is covariant (see \cite{sussQLnum,sussQL}). 

Also, by virtue of the Stewart--Walker gauge invariance lemma \cite{ellisbruni89}, all covariant objects in $S$ that would vanish in the background $\bar S$ (a FLRW cosmology in this case) are gauge invariant (GI), to all orders, and also in the usual sense (as in \cite{bardeen}). The background variables $\mu_*,\,p_*,\,\Theta_*$ do not vanish for $\bar S$, hence they are ``zero order'' GI variables to all orders. The  quantities in LTB spacetimes that vanish for a FLRW cosmology are the tensors $\Pi^{ab},\,\sigma^{ab}$ and $E^{ab}$, given by (\ref{PSEsc}) in terms of the scalar functions $\PP,\,\Sigma$ and $\EE$ in (\ref{PP1})--(\ref{SigEE1}). But from (\ref{Sigma2})--(\ref{EE2}), these functions are basically the fluctuations $\mu-\mu_*$,\, $p-p_*$ and $\Theta-\Theta_*$.  Hence, from (\ref{Da_def}) and (\ref{rad_grads}), the perturbation variables $\Dm,\,\Dp$ and $\Dth$, as well as the gradients $\mu_*',\,p_*'$ and $\Theta_*'$,  are all ``first order'' quantities that are GI to all orders. Therefore, LTB spacetimes in the QL scalar representation $\{A_*,\,\Da\}$ are expressible as spherical, non--linear GIC perturbations on a FLRW background. In the linear limit these perturbations reduce to spherical perturbations in the long wavelength approximation and in the synchronous gauge~\cite{sussQL}.

\begin{figure}[htbp]
\includegraphics[width=4.5in]{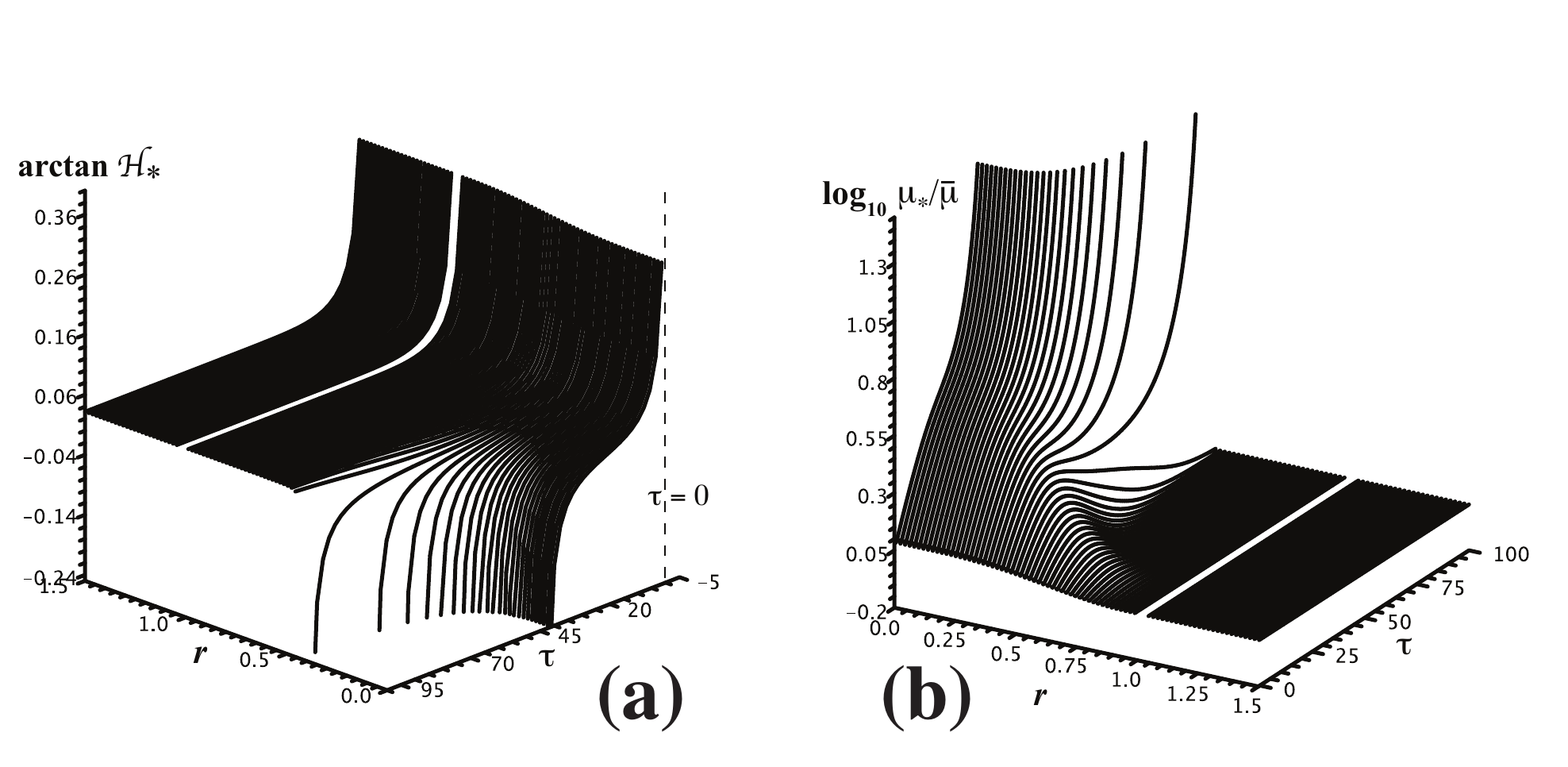}
\caption{{\bf Hubble scalar and density contrast for a Chaplygin gas LTB model.} The figure displays (a) the Hubble scalar and (b) the density contrast for a Chaplygin gas overdensity smoothly matched to a Chaplygin gas FLRW universe (the matching interface $r=r_{\rm b}=1$ is displayed as a white strip).}
\label{fig1}
\end{figure}
\begin{figure}[htbp]
\includegraphics[width=2.5in]{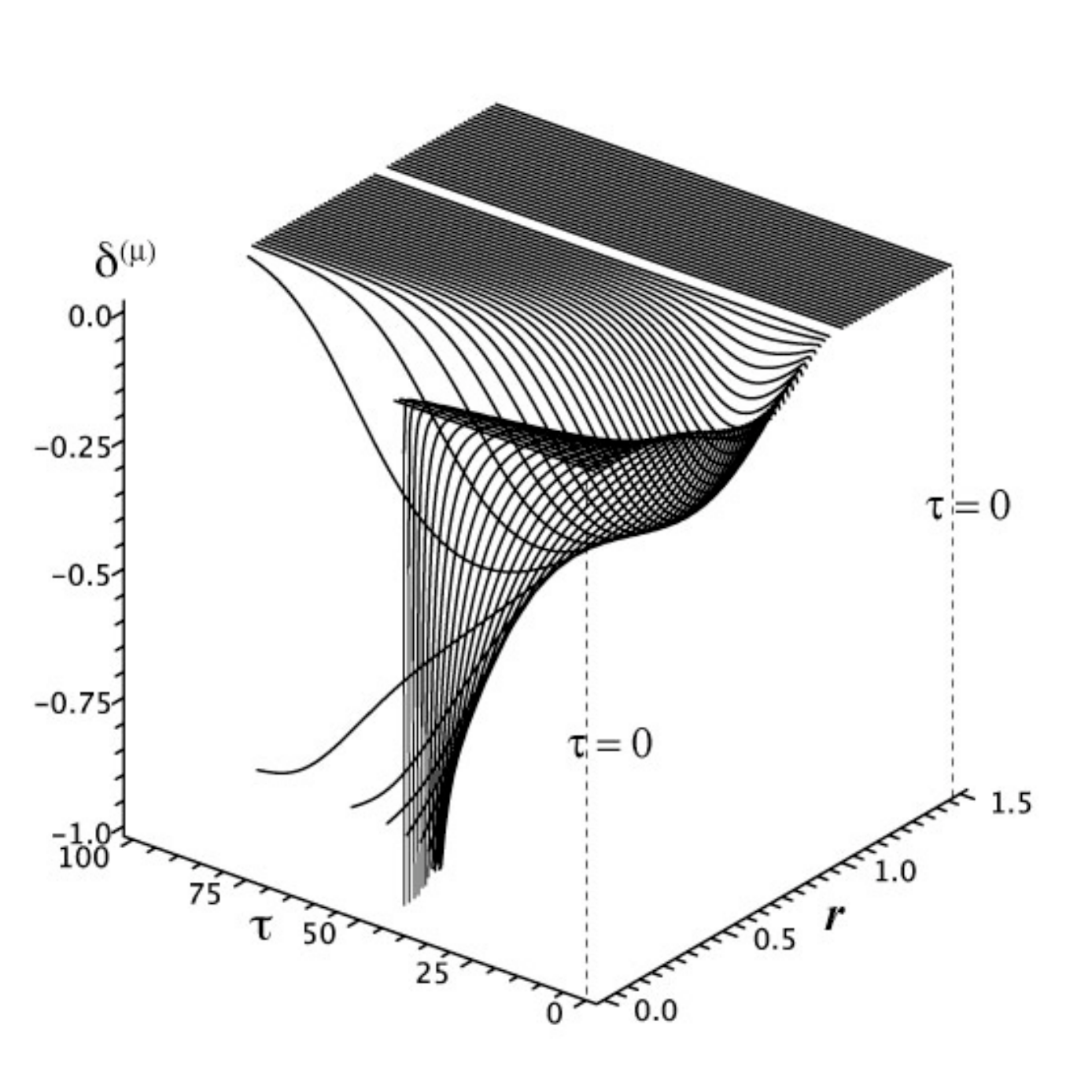}
\caption{{\bf Density perturbations.} The function $\Dm$ for the Chaplygin gas configuration of \ref{fig1}. }
\label{fig3}
\end{figure}

\section{Example: the Chaplygin gas.}

As in any perturbative approach, we need to impose an ``equation of state'' (EOS) between $p_*$ and $\mu_*$ to determine the background subsystem (\ref{evmu_ql})--(\ref{evHH_ql}). Such a choice of an EOS also determines the perturbation equations.   Evidently, the plausibility of the non--linear perturbation formalism must be tested and judged for different EOS according to its predictions.  We consider as an example the Chaplygin gas as a  fully relativistic and inhomogeneous LTB spacetime~\cite{sussQLnum}. This source has been examined under inhomogeneous conditions only in a Newtonian context~\cite{chapinhom}, in terms of non--linear but relativistic approximations~\cite{chapinhom2}, or as a static object~\cite{chapstar}. 

The LTB spacetime associated with the ``standard'' Chaplygin gas EOS~\cite{chaplygin1} is that characterized by the following the following EOS
\begin{equation}p_* = -\frac{\alpha}{\mu_*},\qquad w = -\frac{\alpha}{\mu_*^2},
\label{Chap_eqst}\end{equation}
where $\alpha$ is a constant. Following (\ref{rad_grads}), the relation between pressure and density  fluctuations and perturbations is then
\begin{equation} p-p_* = 
\frac{\alpha}{\mu_*^2}\,(\mu-\mu_*),\qquad
     \Dp = -\Dm.\label{fluc_Chap}\end{equation}
The evolution equations are simply (\ref{evmu_ql})--(\ref{evDth_ql}) specialized for the EOS  (\ref{Chap_eqst}) and with $\Dp$ given by (\ref{fluc_Chap}). Since, from (\ref{Da_def}), we have $p=p_*[1+\Dp]$ and $\mu=\mu_*[1+\Dm]$, equations (\ref{fluc_Chap})  imply the relation
\begin{equation}p = -\frac{\alpha}{\mu}\,[1-(\Dm)^2],\label{Chap_local}
\end{equation}
which can also be interpreted as a first order virial correction to the FLRW EOS containing squared fluctuations that convey the effect of long range interactions. This type of correction is qualitatively analogous to that arising in the ideal gas under a self--gravitating regime in Newtonian systems~\cite{saslaw} (see also \cite{sussQL}). As long as we ignore the fundamental physics of the Chaplygin gas, we cannot rule out the possibility that we might be describing important non--local effects by using the EOS (\ref{Chap_eqst}) and by having the local variables $p,\,\mu$ given by expressions like (\ref{Chap_local}). 

We present here three graphs obtained from the numerical solution of (\ref{evmu_ql})--(\ref{evDth_ql}) for the Chaplygin gas ``top hat'' model, constructed by smoothy matching a section of a Chaplygin gas LTB model ($0\leq r \leq r_b$) with a Chaplygin gas spatially flat FLRW universe ($r>r_b$), with $r_b=1$. As shown in panel (a) of figure \ref{fig1}, the QL Hubble scalar $\HH_*=\Theta_*/3$ passes from infinite values at an initial singularity for all $r$. In the overdensity region we see that $\HH_*\to-\infty$, indicating a collapse to a black hole, while in the FLRW region $\HH_*$ tends to a positive constant that can be identified with the  $\Lambda$ value that FLRW Chaplygin models tend to asymptotically. In panel (b) of figure \ref{fig1}, we plot the ``density contrast'' ratio of the QL density $\mu_*$ to the FLRW density $\bar\mu$. The ratio diverges as inner layers of the overdensity collapse into the black hole while external layers blend into the cosmic background as $r\to 1$. Figure \ref{fig3} displays the ``exact'' non--linear perturbation $\Dm$. This function is negative (overdensity) and close to zero near the center and the matching interface (where radial gradients are small). It is significantly different from zero in the areas where the radial gradient is large and $\Dm\to -1$ as inner layers collapse into a black hole and at the initial singularity for all layers (not shown). A more detailed numerical study of the Chaplygin gas LTB model was undertaken in \cite{sussQLnum}.


\begin{theacknowledgments}
  The author is grateful for the hospitality of the organizing committee of the 3rd International Meeting on Gravitation and Cosmology. 
\end{theacknowledgments}



\bibliographystyle{aipproc}   

\bibliography{sample}

\IfFileExists{\jobname.bbl}{}
 {\typeout{}
  \typeout{******************************************}
  \typeout{** Please run "bibtex \jobname" to optain}
  \typeout{** the bibliography and then re-run LaTeX}
  \typeout{** twice to fix the references!}
  \typeout{******************************************}
  \typeout{}
 }

\end{document}